\documentclass[preprint,authoryear,review,11pt]{elsarticle}

\usepackage{lineno}

 \usepackage{graphics}
 \usepackage{graphicx}
\usepackage[caption = false]{subfig}
\usepackage{amssymb}
\usepackage{amsmath}
\usepackage{float}

\usepackage{chngcntr}
\counterwithout{figure}{section}

\newcommand{\dbh}{\mbox{dbh}}

\begin{document}

\begin{frontmatter}

 \title{Simulated effects of site salinity and inundation on long-term growth trajectory and carbon sequestration in monospecific \textit{Rhizophora mucronata} plantation in the Philippines}
 \author[label 1]{Drandreb Earl O. Juanico\corref{cor1}}
 \author[label 2]{Severino G. Salmo III} 
 \cortext[cor1]{Corresponding author: \texttt{reb.juanico@tip.edu.ph}}
 \address[label 1]{Technological Institute of the Philippines, P. Casal Street, Quiapo 1001, Philippines}
\address[label 2]{Ateneo de Manila University, Loyola Heights, Quezon City 1108, Philippines}

\begin{abstract}
A mathematical model of coastal forest growth is proposed to describe and test the effects of salinity and inundation in the long-term growth performance and carbon sequestration of monospecific mangrove (\textit{Rhizophora mucronata}) plantation in the Philippines. We used allometry in expressing the mangrove growth equation, and stochasticity in scheduling population-level events that drive the development of the mangrove forest. Analysis of the model unveils an index, $\xi$, that could be used in assessing a strategy which could promote optimum carbon-stock accumulation in the long run.
If initial plot is configured such that $\xi > 1$, the \textit{R. mucronata} plantations could achieve an above-ground biomass per hectare (AGB) of $1000\mbox{ t/ha}$, or about $500\mbox{ tC/ha}$, in approximately $200$ to $250$ years post planting. In contrast, the current restoration strategy implemented in the Philippines corresponds to the case that $\xi <1$. Consequently, the restored mangroves could not achieve stable growth without the support of costly human assistance such as frequent replanting. Rather, through that typical strategy and in the absence of assistance, the AGB decreases with time until all trees die. Mangrove restoration could therefore be planned strategically to mitigate costly and wasteful implementation. The proposed index $\xi$ thus serves as an early indicator for the progress or demise of restored mangroves.
\end{abstract}


\end{frontmatter}


\section{Introduction}

Mangroves are known to provide several socio-ecological and ecosystem services such as timber and fisheries production, nutrient regulation, shoreline protection, etc. [see for example a review by~\cite{Lee}]. Unfortunately, mangroves are being lost worldwide at an alarming rate of 1\% per year due to various natural and anthropogenic causes~\citep{FAO2007}. In the Philippines, the total mangrove forest cover decreased by 51.80\% between 1918 and 2010. Particularly, an annual loss rate of 0.52\% from 1990 to 2010 was mainly attributed to aquaculture development~\citep{Long2013}. The depletion of mangroves shall result in the reduction of ecosystem functionality, and may expose coastal areas to higher vulnerability against natural disasters such as typhoons and storm surges~\citep{Duke2007}.

In November 2013, Super Typhoon Haiyan ravaged the Eastern Visayas region in Central Philippines. It is the strongest in historical records that ever made landfall~\citep{Zhang2013}. In the quest for solutions to mitigate similar future coastal disasters, mangroves along coastal fringes are being considered for their potential to protect against storm surges~\citep{Temmerman2013,Schmitt2013}. Mangroves are known to attenuate waves by as much as $75$\% through its vast underground root networks and high vegetation structural complexity, but only if mangroves have a wide extent of at least one km~\citep{McIvor2012}.


Other than coastal protection, mangrove restoration along coastal fringes also contributes toward climate change adaptation and mitigation strategies~\citep{Duarte2013}; for instance, through the sequestration of atmospheric $\mbox{CO}_2$~\citep{Donato,Masera}. Based on field measurements by~\cite{Salmo}, mangrove forests have been found to increase carbon stock over time through the accumulation of biomass as the forest grows and develops. Carbon stock accumulation demonstrates the capacity of mangroves to absorb atmospheric $\mbox{CO}_2$, and thereby contributes in regulating the impacts of global warming.

But in order to promote carbon stock accumulation, the mangrove forest must increase its total area and tree density (i.e., number of trees per unit area). The effectiveness of a restoration strategy in achieving that goal depends on the growth and survival of planted mangroves. In the Philippines, most restored mangroves are monospecific, oftentimes using the genus \textit{Rhizophora}. The initial mangrove plots are situated in sub-optimal positions that are highly saline and too frequently inundated~\citep{SamsonRollon}.

Optimal restoration strategy usually involves the planting of appropriate species at the right locations [see review by~\cite{Primavera2008}]. The authors criticized the wide use of \textit{Rhizophora} seedlings in restoration programs for convenience even if evidence points that \textit{Rhizophora} sp. planted in coastal fringes could not survive in the long run nor attenuate strong wave action. Similarly,~\cite{Salmo2014} argued that since planted mangroves are less diverse, these plantations (even at $\sim\!\!20\mbox{ years}$) offer less resilience against strong typhoon as documented in the damages brought by Typhoon Chan-hom in Lingayen Gulf (northwestern Philippines) in May 2009.

Due to the long periods of time necessary for the growth and development of mangrove forests, assessment of restoration strategies is challenging. Mathematical models can nevertheless circumvent that difficulty~\citep{BergerReview}. Through model simulations, long-term trajectories of mangrove growth and carbon-stock accumulation can be generated. Simulations can be implemented with settings that mimic the conditions of typical restoration sites in the Philippines. 

Here, we propose a mathematical model of mangrove forest growth in order to gain insights on generic features that an optimal restoration strategy must have. By offering a simplified description, the model is made sufficiently tractable upon which mathematical analysis could be performed. Furthermore, it is a substantial innovation from prevailing models of forest dynamics. The biophysical growth is described by a function that considers natural allometry of mangrove trees. Hence, it does not suffer the spurious singularities found in models based on the regression of data in a given sampled site. The present model also considers naturally-observable stochasticity in the scheduling of demographic events such as seedling establishment (birth), and mortality. Most importantly, the model contributes significant insight for determining which restoration strategy is most likely to succeed in the long run. Restored mangroves must flourish with minimal human assistance, such that carbon stock and tree density both rise or stabilize (in contrast to decline) in time.

\section{Materials and Methods}
\subsection{Overview}
\subsubsection{Purpose}
The model is intended to describe and test monospecific mangrove restoration strategies currently being undertaken in the Philippines. Parameterizations of the model are based on Philippine mangrove plantation data reported by~\cite{Salmo}. Focus is particularly given to the genus \emph{Rhizophora}, which is commonly planted in those restorations. Hypothetical strategies are going to be explored through the model. The long-term trajectory of aboveground biomass is particularly considered. 

\subsubsection{State variables and scales}
Three hierarchical levels comprise the model: the individual mangroves, the mangrove forest population, and the seashore environment. Trees are characterized mainly by the state variable referred as \emph{diameter at breast height} ($\dbh$), which is a proxy variable for age. Asynchronous development is considered; hence, $\dbh$ is taken as an age indicator instead of time. The $\dbh$ also determines the height $H$, and crown radius $r_{\mbox{\small crown}}$ through scaling arguments based on natural allometry~\citep{Ong2004}. Each mangrove is further characterized by its fixed position on the seashore, which in turn determines its functional response to stress gradients and to resource competition. 

A forest population refers to the entire collection of mangrove stands over a certain area of interest. Population is characterized by abundance, and the convex hull (i.e., the widest horizontal convex area enclosing the entire population). The population may further be subdivided according to the subpopulations of three main developmental stages: seedling ($
\dbh < 2.5\mbox{ cm}$), sapling ($2.5\leq \dbh < 5\mbox{ cm}$), and tree ($\dbh > 5\mbox{ cm}$). The ratio of population size against the area $A$ of the convex hull is interpreted as population density (scaled into number of seedlings/saplings/trees per hectare). The forest is further characterized by its aboveground biomass (AGB), which is related by allometry to $\dbh$ according to the following equation defined by~\cite{Komiyama}:

\begin{equation}
\mbox{AGB} = 0.235~\dbh^{2.42}\quad\mbox{t/ha}
\label{eq:agb}
\end{equation}

The carbon stock is about half of the AGB~\citep{Masera}. The seashore environment and its stress factors are taken to be the highest level in the hierarchy. Nutrient resources in the environment are assumed to be distributed uniformly in space and remain fixed in time. Two stressors relevant to mangrove survival are here considered: salinity and inundation. The salinity and inundation fields increase with seaward gradient. However, the stress fields are assumed not to vary with time. Individual mangroves subjected to stressors will have a general slowdown of growth. The response is nonlinear with respect to salinity, and linear with inundation. Lastly, competition between individuals depends on $\dbh$ and the relative position of an individual in the patch. This method  is adapted from the \emph{Field of Neighborhood} (FON) approach originally proposed by~\citet{BergerHildenbrandt}.

\subsubsection{Process overview and scheduling}
\begin{figure}
\centering
\includegraphics[scale=.17]{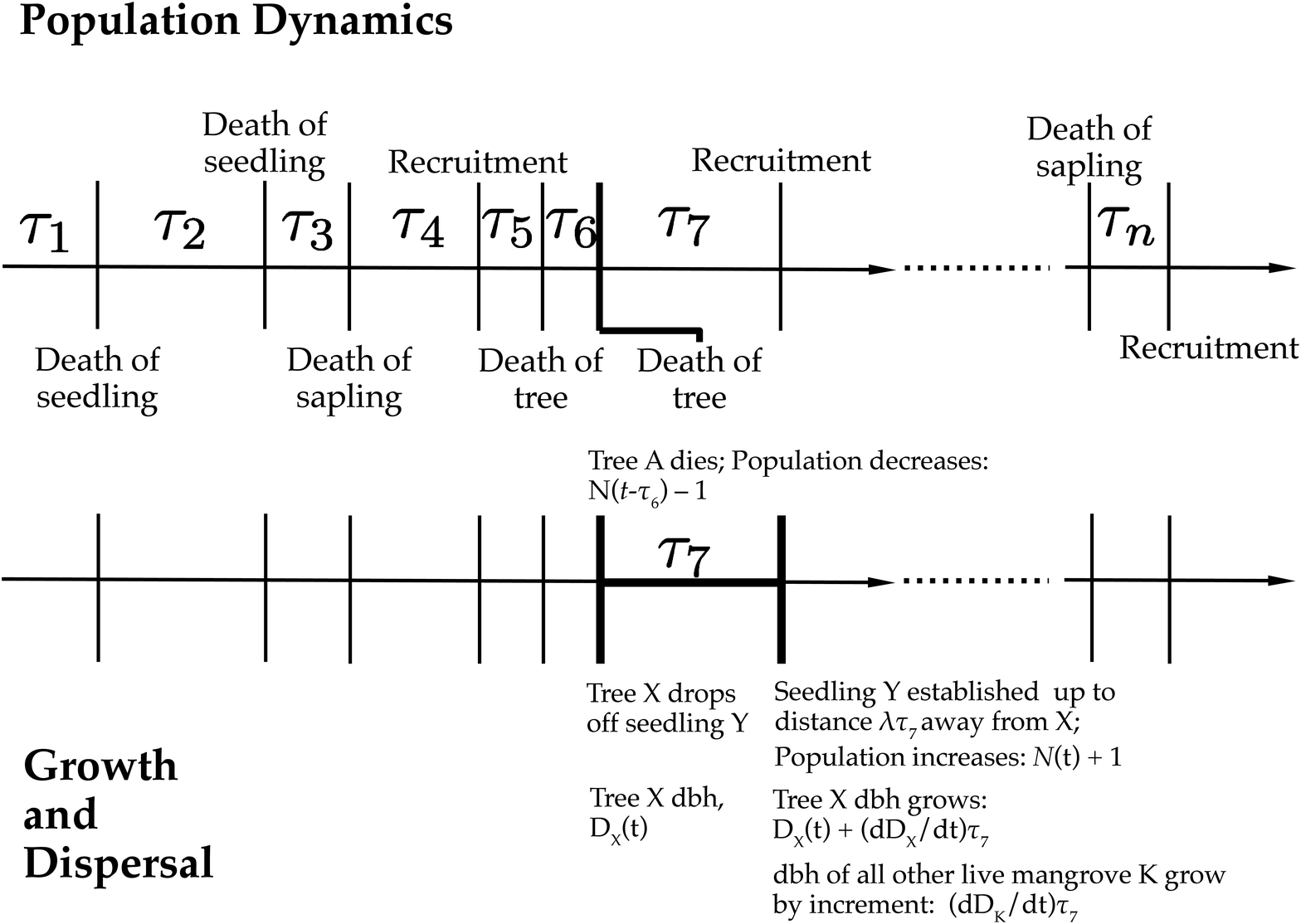}
\caption{Representative time course of a simulation. The $n$-th iteration occurs across a time step $\tau_n$. The sequence $\left\{\tau_n\right\}$ is a random sequence generated by the waiting time distribution of a Poisson process. At the end of each iteration, the population size $N$ changes by a unit due to any of the possible demographic events: recruitment or mortality. The choice of event is random. The time step is assumed short enough so that only one event takes place within it.  Meanwhile, individual growth proceeds during an iteration according to the dbh increment $dD/dt$ multiplied by $\tau_n$. Propagule dispersal during an iteration is modelled as a random walk in an annular region, the outer radius of which is the product of a dispersal rate and the time interval $\tau_n$. At the end of the iteration, the propagule establishes and becomes a seedling with $\dbh=0.5\mbox{ cm}$.}
\label{fig:schedule}
\end{figure}

The model simulates daily time intervals (i.e., $dt = 1\mbox{ day}$). But the length of time step $\tau$ per iteration varies randomly, in accordance with a Poisson process (see Figure~\ref{fig:schedule}).  Across each iteration, a randomly chosen tree gives rise to one recruit, or a randomly selected living mangrove dies with a probability related to its developmental stage. The maximum isotropic displacement of the recruit is proportional to $\tau$. Meanwhile, over an iteration, all living mangroves grow proportionately with $\tau$.  

\subsection{Design concepts}
\subsubsection{Emergence}
A spatial forest pattern emerges through the asynchronous seedling establishment and growth, the competitive interaction between individuals, and through the constraints imposed by the stressors. The mangrove life cycle is entirely represented by empirical rules describing mortality and dispersal as probabilities, whereas $\dbh$ growth is deterministic. While fitness seeking is not modelled explicitly, the competition and stressor fields favor growth and survival in some locations of the forest over other locations. 

\subsubsection{Sensing}
Individual mangroves are assumed to sense and respond to the growth constraints imposed by competition and stressors in its immediate surroundings. In particular, the individual mangrove ``perceives" the aggregate effect of the presence of other nearby mangroves.

Mangrove age is not clocked with time, but rather through its developmental stage in terms of its $\dbh$. Thus, a mangrove could linger in a stage longer or shorter than the average depending on the extent of growth constraints to which it is subjected at any given moment. For example, when a mangrove's $\dbh$ is above a certain empirically known threshold, the mangrove identifies its maturity as a tree and thus only then will initiate seedling production. 

\subsubsection{Interaction}
Mangroves interact with one another at a distance from their trunk axis. A ``force-like field" has been proposed by~\cite{BergerHildenbrandt} to describe such interaction. The field decays exponentially with distance from the trunk axis. The existence of the field may be justified by the presence of an extended root system below ground, and crown system above the trunk. Competition ensues from the field interactions: root systems compete for belowground spaces for anchorage, and crown systems compete for sunlight. Competition slows down the growth of mangrove by a factor that is a function of the strength of the aggregate field it encounters from other mangroves. As a consequence of the slow growth, a mangrove could have a higher probability of dying if its size remains small for a longer time as compared to its neighbors with larger sizes. 

\subsubsection{Stochasticity}
\begin{figure}
\centering
\includegraphics[scale=.25]{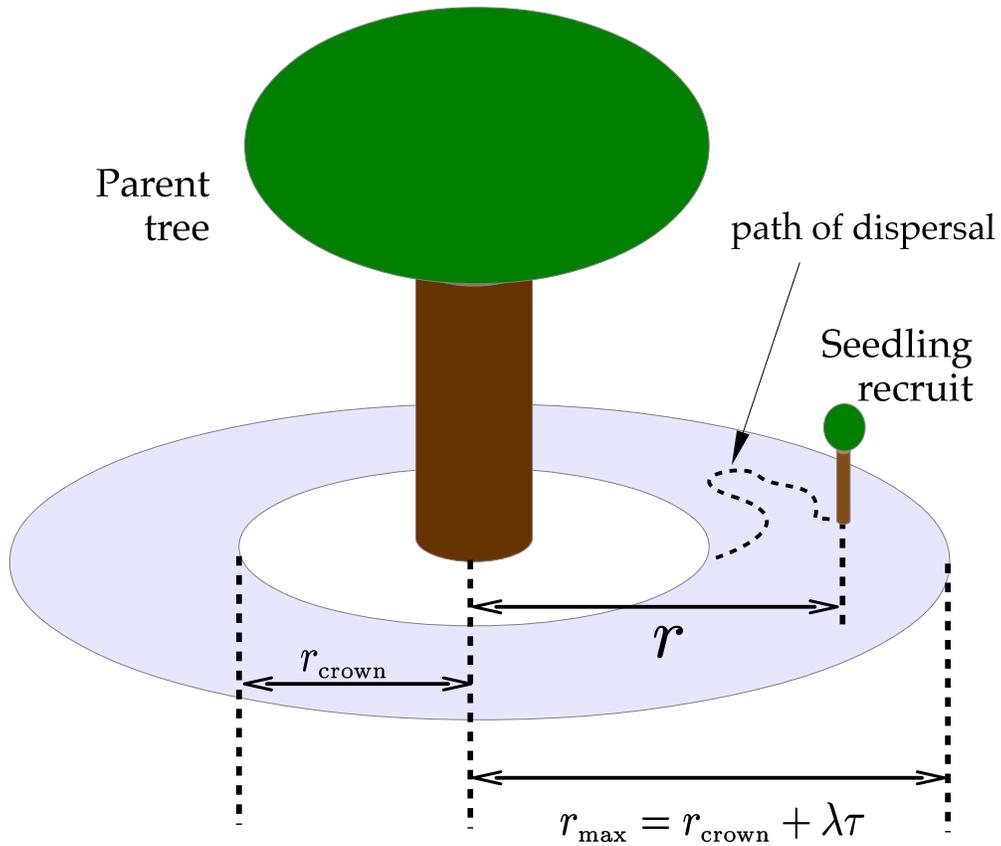}
\caption{Radial dispersal of a propagule leading to a seedling recruit. A propagule from a parent tree disperses along a Brownian-like path from any point along the rim of a circle with the crown radius, $r_{\mbox{\small crown}}$. The dispersal radius $r$ of the propagule before it establishes at a fixed position is the radial horizontal length from the vertical axis of the parent tree. The maximum possible displacement over a time interval $\tau$ is $r_{\mbox{\small max}}=r_{\mbox{\small crown}}+\lambda\tau$ where $\lambda$ is an empirically known dispersal rate. In general, $r \leq r_{\mbox{max}}$. Dispersal is assumed to be isotropic so that the seedling could establish at any direction around the parent tree as long as there is available space.}.
\label{fig:dispersal}
\end{figure}

Randomness plays an essential role in the model. The changes in population size are categorized between recruitment and mortality. A time unit of $1$ day is assumed short enough so that a maximum of only one unit of change in the population size could possibly occur within it. As to which population size-changing event (among those indicated in Fig.~\ref{fig:schedule}) will occur within a time unit, the choice is randomly determined. Also due to stochasticity, an event is not always guaranteed to happen at every time unit. The focal individual is also determined at random. In a recruitment event, the seedling is randomly placed within an annular region surrounding the parent tree, as shown in Figure~\ref{fig:dispersal}.

The time course is also stochastic such that the interval between unit changes in the population is random in length, as depicted in Figure~\ref{fig:schedule}. Hence, the population dynamics of the forest can be interpreted as a Poisson process, which is a common model for birth-death processes~\citep{Gardiner}.

\subsection{Details}

\subsubsection{Initialization}
\begin{figure}
\centering
\includegraphics[scale=.25]{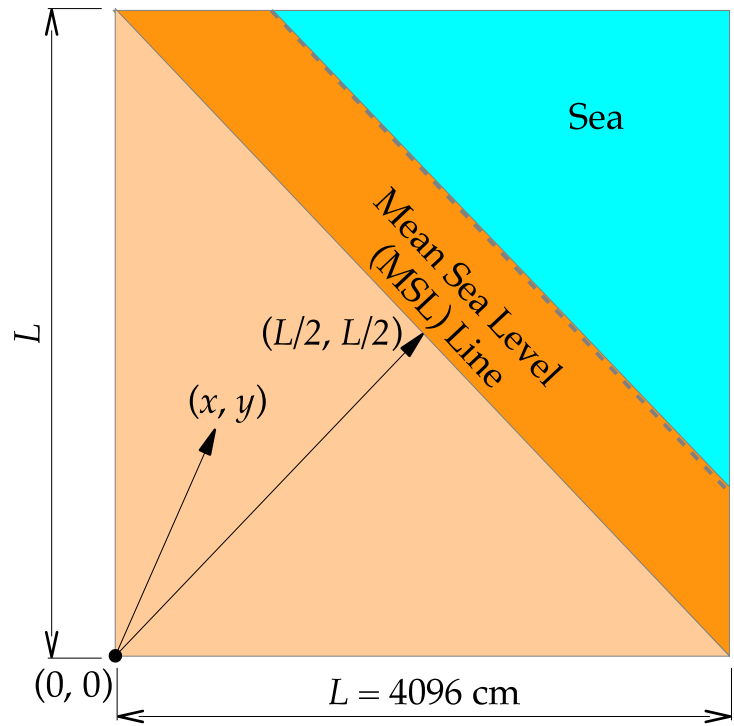}
\caption{Top view of the simulated seashore spanning the middle and part of the upper intertidal zone. The mean sea level (MSL) delineates the average coastline. The MSL line separates the shore below the MSL (dark shade) and above it (light shade). The dashed line represents the lowest edge of the middle intertidal just above the lower intertidal. The lower intertidal zone just beyond the dashed boundary may be exposed during low water spring tide. The origin $(0,0)$ of the coordinate system is located at the bottom left corner. Any point on the coordinate system is described by coordinates $(x,y)$, where $x$ and $y$ are the horizontal and vertical distance from origin, respectively. Environment fields are defined as functions of these coordinates. The direction of increasing elevation is toward $(0,0)$.}
\label{fig:grid}
\end{figure}

A seashore spanning the middle and part of the upper intertidal zone is simulated as a patch having the simple geometry shown in Figure~\ref{fig:grid}. A point on the patch is described by its coordinates $(x,y)$. Boundaries are closed so that forest growth dynamics are only defined within the patch perimeter.

The seashore is initially occupied by seedlings with $\dbh = 0.5\mbox{ cm}$. Positions of individual seedlings are randomly chosen within a designated initial plot on the seashore. Seedlings could not overlap. The seedling density distributed across the initial plot is fixed. Different stochastic realizations of the model are performed for a given initial plot configuration.

\subparagraph*{Salinity field.}  A salinity field $S(x,y)$ is attributed to the patch shown in Fig.~\ref{fig:grid}, such that $S(0,0)=0$. The porewater salinity is assumed to be higher at locations nearer to the sea as opposed to the location closer to $(0,0)$. At sea, the salinity stabilizes to about $72\mbox{ ppt}$, which is rather extreme but here considered for the purpose of subjecting mangroves to poor growth (i.e., high salinity) conditions. The salinity field is thus defined as follows:
\begin{equation}
S(x,y) = \min\left[\frac{72}{L}\left(x+y\right), 72\right]
\label{eq:salinity_field}
\end{equation}
where $L=4096\mbox{ cm}$ as depicted in Fig.~\ref{fig:grid}. Consequently, the gradient of the salinity field is about $1.76\mbox{ ppt/m}$.

\subparagraph*{Inundation field.} The tidal inundation field $I(x,y)$ is related to the elevation profile of the patch. The elevation profile is such that $I(0,0)=0$; in other words, the point is high enough that no tide could reach it all year round. The field is defined as follows,
\begin{equation}
I(x,y) = \min\left[\frac{0.8}{L}\left(x+y\right),1\right]
\label{eq:inundation_field}
\end{equation}
where $0.8/L$ corresponds to the slope of the coast of approximately $2\mbox{ cm}$ decrease in elevation for every meter of advance toward the sea, which implies that the seashore advances some 25\% further at low tide (as depicted by the dashed line in Fig.~\ref{fig:grid}). This is consistent with the average tidal fluctuation of about $2\mbox{ m}$ for Philippine coastlines. Beyond the low tide line (dashed line in Fig.~\ref{fig:grid}), land is persistently submerged (except possibly during low water spring tide); hence, $I=1$ beyond the middle intertidal zone. 

\subsubsection{Input}
The model requires the initial number of seedlings, and the position and configuration of the initial plot. An individual mangrove responds to stressor gradients with an effective reduction on their growth rates. The salinity and inundation response of a mangrove are denoted as $\sigma = \sigma[S(x,y)] = \sigma(x,y)$, and $\eta = \eta[I(x,y)] = \eta(x,y)$, respectively.

\subparagraph*{Salinity response. } We apply the salinity response proposed by~\cite{BergerHildenbrandt}. Hence, $\sigma(x,y)\in [0,1]$ is defined as a logistic decay function of salinity $S(x,y)$, as follows:
\begin{equation}
\sigma(x,y) = \left[1+\exp\left(\frac{S(x,y)-\tilde{S}}{\Delta S}\right)\right]^{-1}
\label{eq:salinity_response}
\end{equation}
where $\tilde{S}$ is the species specific critical salinity level above which a plant is not likely to grow and survive. $\Delta S$ is a species-specific tolerance to salinity, which indicates a range of salinity levels about  $S=\tilde{S}$ for which the response undergoes abrupt transition. In the case of \emph{R. mucronata}, we set $\tilde{S} = 72\mbox{ ppt}$ and $\Delta S = 4\mbox{ ppt}$ in accordance with parameter settings for the genus \textit{Rhizophora} by~\cite{BergerHildenbrandt}.

\subparagraph*{Inundation response.} A mangrove is expected to grow faster if it is subjected to shorter inundation times (or being above water most of the time). Thus, the inundation response $\eta(x,y)\in [0,1]$ is defined by the following function,
\begin{equation}
\eta(x,y) = 1-I(x,y)
\label{eq:inundation_response}
\end{equation}

\subsubsection{Submodels}

\subparagraph*{Competition field.} Plant competition has been made spatially explicit in the FON approach~\citep{BergerHildenbrandt}. Mangroves compete with each other via field interaction. Each individual mangrove is set to give rise to a rotationally symmetric scalar field which decays with radial distance from its trunk axis. Suppose a mangrove is located such that its trunk axis stands on $(x_0,y_0)$ of the horizontal plane. Let $(x,y)$ be the field coordinate so that the radial distance with respect to $(x_0,y_0)$ is $r=\sqrt{\left(x-x_0\right)^2+\left(y-y_0\right)^2}$. Specifically, the field $f_0(r)$ (as a function of $r$) generated by the mangrove at position $(x_0,y_0)$ is defined as follows:
\begin{equation}
f_0(r) = 
\begin{cases}
1, & 0\leq r < \dfrac{D}{2}\\
\exp\left[-c\left(r-\dfrac{D}{2}\right)\right], & \dfrac{D}{2}\leq r < r_{\mbox{\small crown}}\\
0, & r > r_{\mbox{\small crown}} 
\end{cases}
\label{eq:scalar_field}
\end{equation}
where $D = \dbh$ of the focal mangrove, and $r_{\mbox{\small crown}}$ is its crown radius related by allometry to $D$~\citep{BergerHildenbrandt}, as follows:
\begin{equation}
r_{\mbox{\small crown}}= \frac{1}{2}\left[22.2\,D^{0.654}\right]\;\mbox{cm}
\label{eq:crown_radius}
\end{equation}
Note that the parameter $c$ in Eq.~(\ref{eq:scalar_field}) acts as a decay constant which controls the strength of the field beyond the trunk axis. The smaller its value, the more extensive is the field; hence, the longer the interaction range of its generated field. On the contrary, a larger $c$ implies more localized fields; hence, weaker long-range interactions between individuals. A value of $c=0$ means that the crowns are strictly rigid. For the simulation results, the value used is $c=0.1$.

A zone of influence with radius $r_{\mbox{\small crown}}$ denotes a territory around a mangrove wherein it ``perceives" the mean field due to all other mangroves in the forest~\citep{BergerHildenbrandtGrimm}. In Fig.~\ref{fig:dispersal}, the said zone is shown as the inner circle with radius $r_{\mbox{\small crown}}$. The aggregate strength of competition encountered by a focal mangrove is determined by superposition of fields due to other mangroves, which intersect with the focal mangrove's zone of influence. Let this sum be denoted as $F(x,y)$ for a focal mangrove located at $(x,y)$. The individual response to competition, $K = K[F(x,y)] = K(x,y)\in [0,1]$, is defined by the following:
\begin{equation}
K(x,y) = 1-2F(x,y)
\label{eq:competition_response}
\end{equation}
wherein $K$ contributes further to the reduction of growth of each individual mangrove.

\subparagraph*{Individual growth rate.} It is assumed that the relationship of height $H$ and leaf area $La$ to $D=\dbh$ can be described in functional form, $H(D)$ and $La(D)$, respectively. A reasonable boundary condition associated with these functions are: $H(0)=0$ and $La(0)=0$, which necessarily mean that in the absence of a trunk, there is no plant.

Forest gap models, which originated with~\cite{Botkin}, emphasize a relationship between $H$ and $D$. This relationship, denoted as $H(D)$, is determined phenomenologically from the nonlinear fitting of available data, which usually consist of a nonzero minimum for $D$. The fundamental flaw in such an imposed relationship is that it does not satisfy the boundary conditions, e.g., $H(0) \neq 0$.

We thus assume that the functional relationships between plant morphological features are expressed by allometry: height, $H(D) = aD^{\alpha}$; and leaf area, $La(D) = bD^{\beta}$. Not only do these functions satisfy the boundary conditions, but also their use has underlying foundations in physiology~\citep{Shingleton}. Allometric scaling is widely applied in forestry literature as reviewed by~\cite{Komiyama}. In expressing the rate of change $dD/dt$ based on these allometric relations, there is no need to assume that $dH/dD=0$ at $D=D_{\mbox{\small max}}$. The underlying problem with such assumption is that $D_{\mbox{\small max}}$ is usually defined empirically through site sampling~\citep{Botkin2011}, rather than as a true biophysical limit. In fact, the existence of maximum height $H_{\mbox{\small max}}\approx 130\mbox{ m}$ for trees on Earth, regardless of species, has sufficient basis on the physics of vertical water transport~\citep{Koch}. $D_{\mbox{\small max}}$ should have a corresponding limit as well, but it is difficult to impose a value for it on the basis of site sampling alone. 

Further advantage of using allometry to relate $H$ and $D$ is that the morphologic parameters in the differential equation $dD/dt = f\left(D;D_{\mbox{\small max}}\right)$ are strictly species specific only. In gap models that impose variants of the $H(D)$ function [originally proposed by~\cite{Botkin}], the resulting $dD/dt$ are implicitly site specific as well. A $D_{\mbox{\small max}}$ value set beforehand from a previous sample is strictly not applicable for a new site even for the same species. If in the new site some trees, for instance, have $D>D_{\mbox{\small max}}$, then the entire model would break down. Such problem arises from the existence of zeroes in the quadratic denominator of $f\left(D;D_{\mbox{\small max}}\right)$ for $D>D_{\mbox{\small max}}$. Hence, substantial revisions to the model parameters are required each time a new site is considered, contrary to previous claims that those parameters are merely species specific~\citep{Botkin2011}. The preceding argument precludes the universality of any model based on non-allometric $H(D)$ relationships~\citep{Botkin2011, Lindner, Risch}.

Generally, forests are characterized by non-uniform spatial gradients. It is assumed that these gradients do not intrinsically affect the allometric exponents. Those exponents are assumed to be universal for any given species growing in a particular climate. Consequently, the value of the exponent could be estimated from aggregated data of plants growing in different zones of a forest. Thus, even if resources are not evenly distributed across the forest, the allometric exponent should remain generally valid because the stunted plants would not only be thinner but also shorter. Therefore, the impact of stressors appears merely as correction factors that effectively slow down the growth rate as expressed by $D' \equiv dD/dt$~\citep{BergerReview}.

By putting together the relevant aspects which hypothetically determine a mangrove's growth rate, the following logistic differential equation expresses the time increment of $\dbh = D$:
\begin{equation}
\frac{d D(x,y,t)}{d t} = \left(\frac{\Omega}{2+\alpha}\right) D^{\beta-\alpha-1}\left[1-\frac{1}{\gamma}\left(\frac{D}{D_{\mbox{\small max}}}\right)^{1+\alpha}\right]\sigma(x,y)\eta(x,y) K(x,y)
\label{eq:dD_dt}
\end{equation}
where $D_{\mbox{\small max}}$ is calculated from the allometric equation: $H_{\mbox{\small max}}=H(D_{\mbox{\small max}}) = aD_{\mbox{\small max}}^{\alpha}$, where $H_{\mbox{\small max}}  = 130\mbox{ m}$~\citep{Koch}, and given the value of $a$ and $\alpha$ for a particular species. Moreover, $0<\gamma\leq 1$ is interpreted as a site-dependent factor such that $\gamma^{1/(1+\alpha)}D_{\mbox{\small max}}$ is the maximum observable $\dbh$ in any given site. While there is no attempt to explicitly relate $\gamma$ to salinity, inundation, storm frequency, resource availability and other environmental factors, its value can be determined empirically. Meanwhile, the parameters $\alpha$ and $\beta$ are strictly species specific, and can be evaluated from empirical data for a certain species even if samples originate from different sites. Based on known allometric data on height and leaf area index, the values $\alpha = 0.95$ and $\beta=2$ are used for \textit{R. mucronata}. Lastly, $\Omega$ is a dimensional scaling parameter that converts the units on the right-hand side of Eq.~(\ref{eq:dD_dt}) into $\mbox{cm}\,\mbox{day}^{-1}$ to be consistent with the units on the left-hand side. We used $\Omega=0.25$ without loss of generality.

\subparagraph*{Recruitment.} Although mangroves generally produce propagules throughout their lifetime, not all of these would successfully establish. In describing recruitment, a rate $k_0$ of successful seedling establishment per tree is defined. Within a random time interval $\tau$, one seedling of a randomly selected parent tree would successfully establish a distance away from that tree due to propagule dispersal, as illustrated in Fig.~\ref{fig:dispersal}.

\subparagraph*{Propagule dispersal.} Observations indicate that hydrochorous seedlings can establish a distance away from the parent tree~\citep{Sousa}. Mangrove seedlings are especially mobile because of their capacity to float and be carried by coastal current. Hydrochorous dispersal of propagules has also been found to be wind assisted~\citep{Stocken2013}. Based on empirical measurements of propagule dispersal rates~\citep{Sousa}, a diffusion rate $\lambda$ is defined. When a tree produces a propagule, that propagule may establish and grow anywhere within an annular region as illustrated in Fig.~\ref{fig:dispersal}. The region has inner radius equal to the parent tree's crown radius, $r_{\mbox{\small crown}}$, as defined in Eq.~(\ref{eq:crown_radius}).  

On the other hand, the outer radius $r_{\mbox{\small max}}$ of the annulus is determined by the time interval $\tau$ before the propagule establishes, and the dispersal rate $\lambda$. A subroutine checks that the position of the established seedling does not overlap with the trunk area of any existing mangrove. For \textit{R. mucronata}, we use $\lambda = 26.67\mbox{ cm day}^{-1}$ deduced from measurements by~\cite{Sousa}.

\subparagraph*{Mortality.} Mangroves are expected to have Type III survivorship curves as any plant in general~\citep{Schaal}. The mortality rate is therefore highest for seedlings and lowest for trees. Sapling mortality rate is in between those rates. In particular, let $k_1$, $k_2$, and $k_3$ be the seedling, sapling, and tree mortality rate, respectively. Hence, $k_1 > k_2 > k_3$ in accordance with Type III survivorship curves.

Treating mortality as a Poisson process, the death of any mangrove in the forest is probabilistic. Within a randomly determined time interval $\tau$, a randomly chosen mangrove dies. Unlike the approach of~\cite{BergerHildenbrandt}, the mortality event is in no way correlated with the growth rate. Death by weather disturbances is, however, not yet considered in the present model.

\subparagraph*{Stochastic population dynamics.} Let $M$, $S_p$, and $S_d$ represent an individual tree, sapling, and seedling, respectively. The demographic events can be further expressed as reaction processes:
\begin{eqnarray}
\mbox{Recruitment:} & M &\overset{k_0}\longrightarrow  M + S_d \label{eq:recruitment}\\
\mbox{Seedling death:} &  S_d &\overset{k_1}\longrightarrow   \varnothing \label{eq:seedling_death}\\
\mbox{Sapling death:} & S_p &\overset{k_2}\longrightarrow \varnothing \label{eq:sapling_death}\\
\mbox{Tree death:} & M &\overset{k_3}\longrightarrow\varnothing\label{eq:tree_death}
\end{eqnarray}

Equations~(\ref{eq:recruitment}) to~(\ref{eq:tree_death}) can be written in terms of a master equation of the probability density $P = P\left(N_M,N_{S_p},N_{S_d},t\right)$. With the assumption that the population size is $N = N_M + N_{S_p} + N_{S_d} \gg 1$, the master equation yields as a linear approximation the following deterministic system:

\begin{equation}
\begin{bmatrix}
m' \\ s_p' \\ s_d'
\end{bmatrix} \approx \begin{bmatrix}
-k_3 & \bar{D}_5' & 0\\
0 & -k_2 & \bar{D}_{2.5}' \\
k_0 & 0  & -k_1
\end{bmatrix} \begin{bmatrix}
m \\ s_p \\ s_d
\end{bmatrix}
\label{eq:ODEsystem}
\end{equation}
where $m = N_M/A$, $s_p = N_{S_p}/A$, and $s_d = N_{S_d}/A$ (given $A$ is the convex-hull area), and which has the trivial fixed point, $\Gamma^* \equiv \left(m,s_p,s_d\right) = (0,0,0)$. The primes denote time derivatives, and $\bar{D}'$ is a sample average of Eq.~(\ref{eq:dD_dt}).  Specifically, $\bar{D}_5'=D'(D=5\mbox{ cm})$ and $\bar{D}_{2.5}'=D'(D=2.5\mbox{ cm})$ denote the mean transition rates from sapling to tree, and from seedling to sapling, respectively. 

The value of $\bar{D}'$ could ultimately be computed with respect to the spatial variables, but such an attempt shall be reported in another study. Rather, it suffices to show from the eigenvalues of the square matrix in Eq.~(\ref{eq:ODEsystem}) whether the trivial fixed point, $\Gamma^*$, is either stable or unstable. In particular, we consider $k_0 > 0$ and all other demographic rates are proportional to $k_0$. Barring any human assistance or extreme weather disturbances, a stable $\Gamma^*$ makes forest collapse inevitable, whereas an unstable one implies that the tree population can only increase in time. The latter case is desirable as a dense greenbelt along the coast is preferred for promoting carbon stock accumulation.

\subsection{Simulation Experiments}

\subparagraph*{Gillespie's direct method.} The reaction processes associated with the demographic events denoted in Equations~(\ref{eq:recruitment}) to~(\ref{eq:tree_death}) can be written down as propensities, defined in the following: 
\begin{eqnarray}
\mbox{Recruitment:} & T\left[N_M,N_{S_p},N_{S_d}\rightarrow N_{S_d}+1\right] = k_0 N_M  \label{eq:propensity_recruitment}\\
\mbox{Seedling death:} & T\left[N_M,N_{S_p},N_{S_d}\rightarrow N_{S_d}-1\right] = k_1 N_{S_d}  \label{eq:propensity_seedling_death}\\
\mbox{Sapling death:} & T\left[N_M,S_p\rightarrow N_{S_p}-1,N_{S_d}\right] = k_2N_{S_p}  \label{eq:propensity_sapling_death}\\
\mbox{Tree death:} & T\left[N_M\rightarrow N_M-1,N_{S_p},N_{S_d}\right] = k_3 N_M \label{eq:propensity_tree_death}
\end{eqnarray}

The processes associated with the life-stage transitions may also be expressed in a similar manner: 
\begin{eqnarray}
\mbox{$S_p \rightarrow M$:} & T\left[N_M\rightarrow N_M+1,N_{S_p}\rightarrow N_{S_p}-1,N_{S_d}\right] = \tilde{D}'_5 N_{S_p}\\
\mbox{$S_d \rightarrow S_p$:} & T\left[N_M,S_p\rightarrow N_{S_p}+1,N_{S_d}\rightarrow N_{S_d}-1\right] = \tilde{D}'_{2.5} N_{S_d}
\end{eqnarray}

However, these are only implicitly included in the Gillespie simulations through the change induced by growth, as described by Eq.~(\ref{eq:dD_dt}). Using the propensities defined in Eqs.~(\ref{eq:propensity_recruitment}) to~(\ref{eq:propensity_tree_death}), standard procedures of carrying out the Gillespie algorithm, detailed elsewhere~\citep{Gillespie}, are applied.

\section{Results}

\subsection{Initial configuration of mangrove forest}
\begin{table}
\centering
\caption{Demographic rates used in the simulations without sufficient loss of generality. The settings are consistent with Type III survivorship curves~\citep{Schaal},  attributed to plants like mangroves (i.e., $k_1 > k_2 > k_3$).}
\begin{tabular}{c|c|c}\hline
\textbf{Demographic rate} & \textbf{Symbol} & \textbf{Value}\\\hline
Seedling establishment rate & $k_0$ &   $\dfrac{1}{3650}\mbox{ per day per tree}$\\\hline
Seedling mortality rate & $k_1$ & $2k_0\mbox{ per day per seedling}$\\\hline
Sapling mortality rate & $k_2$ & $k_0\mbox{ per day per sapling}$\\\hline
Tree mortality rate & $k_3$ & $\dfrac{5}{6}k_0\mbox{ per day per tree}$\\\hline
\end{tabular}
\label{table:demographic-rates}
\end{table}

\begin{figure}
\centering
\subfloat[Typical plot: $0.8L\rightarrow 1.1L$]
{\includegraphics[scale=.37]{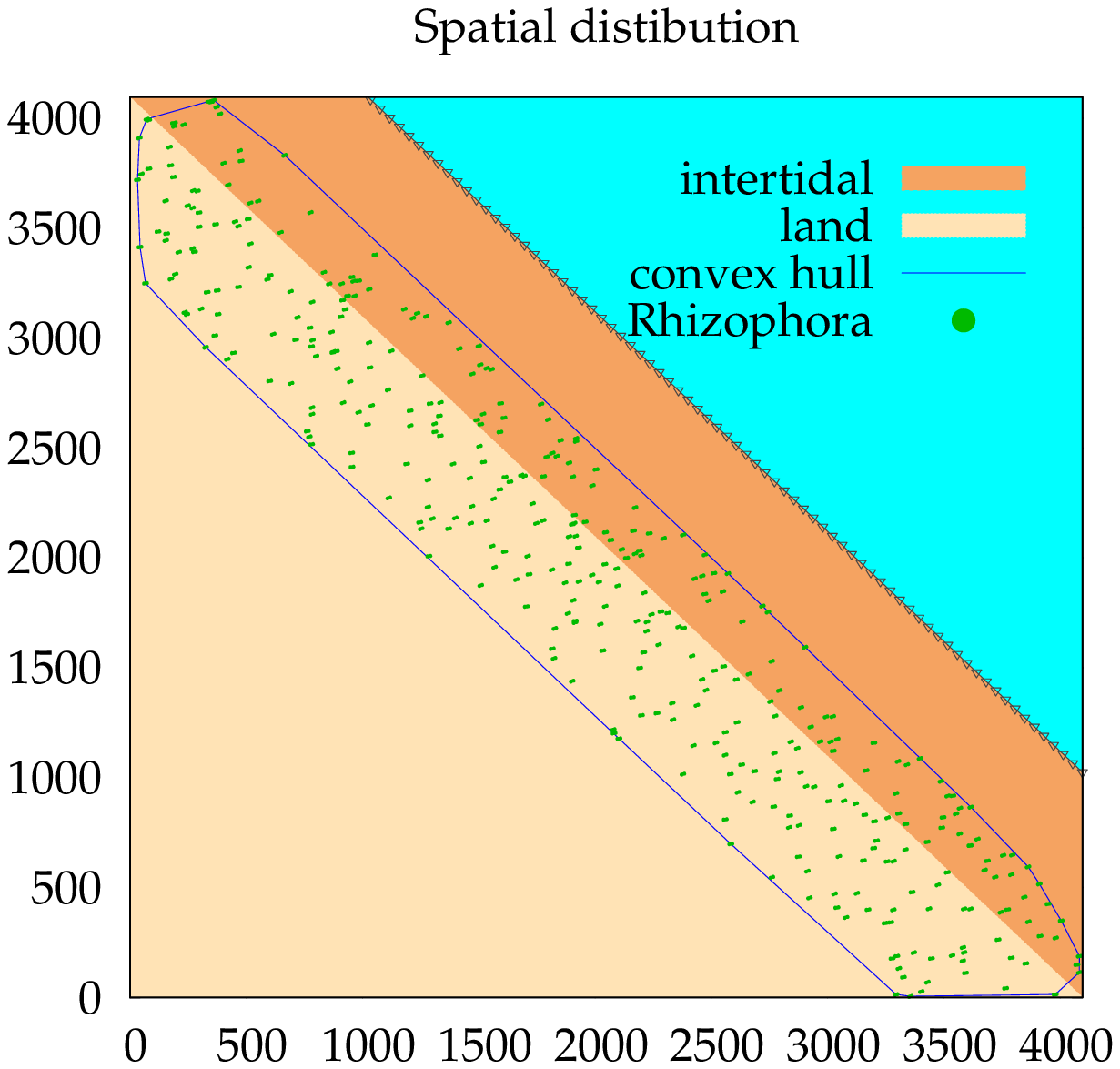}} 
\subfloat[Above-MSL plot: $0.6L\!\rightarrow \!0.8L$]
{\includegraphics[scale=.37]{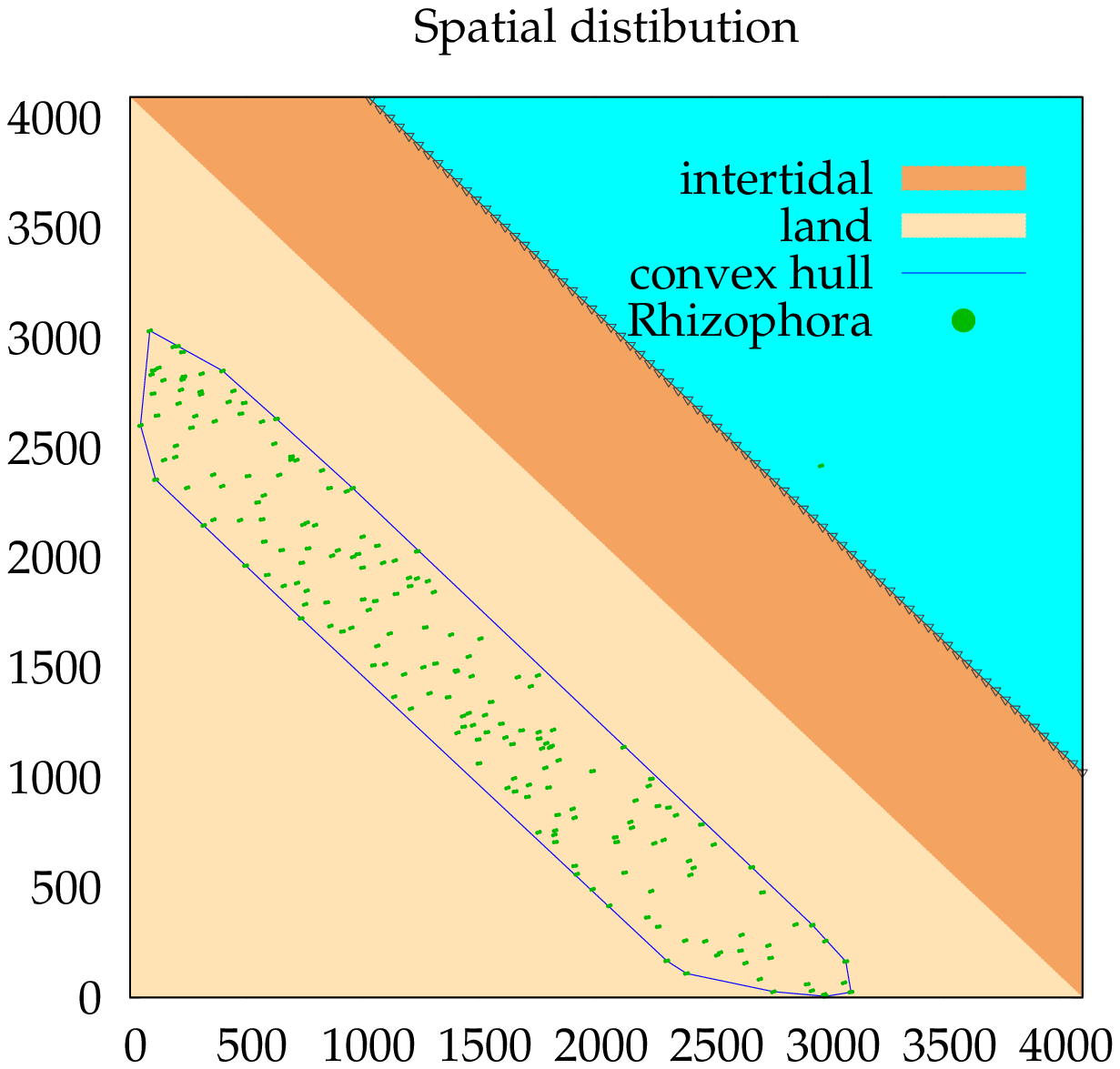}}
\subfloat[Above MSL: $0.4L\!\rightarrow\!0.6L$ ]
{\includegraphics[scale=.37]{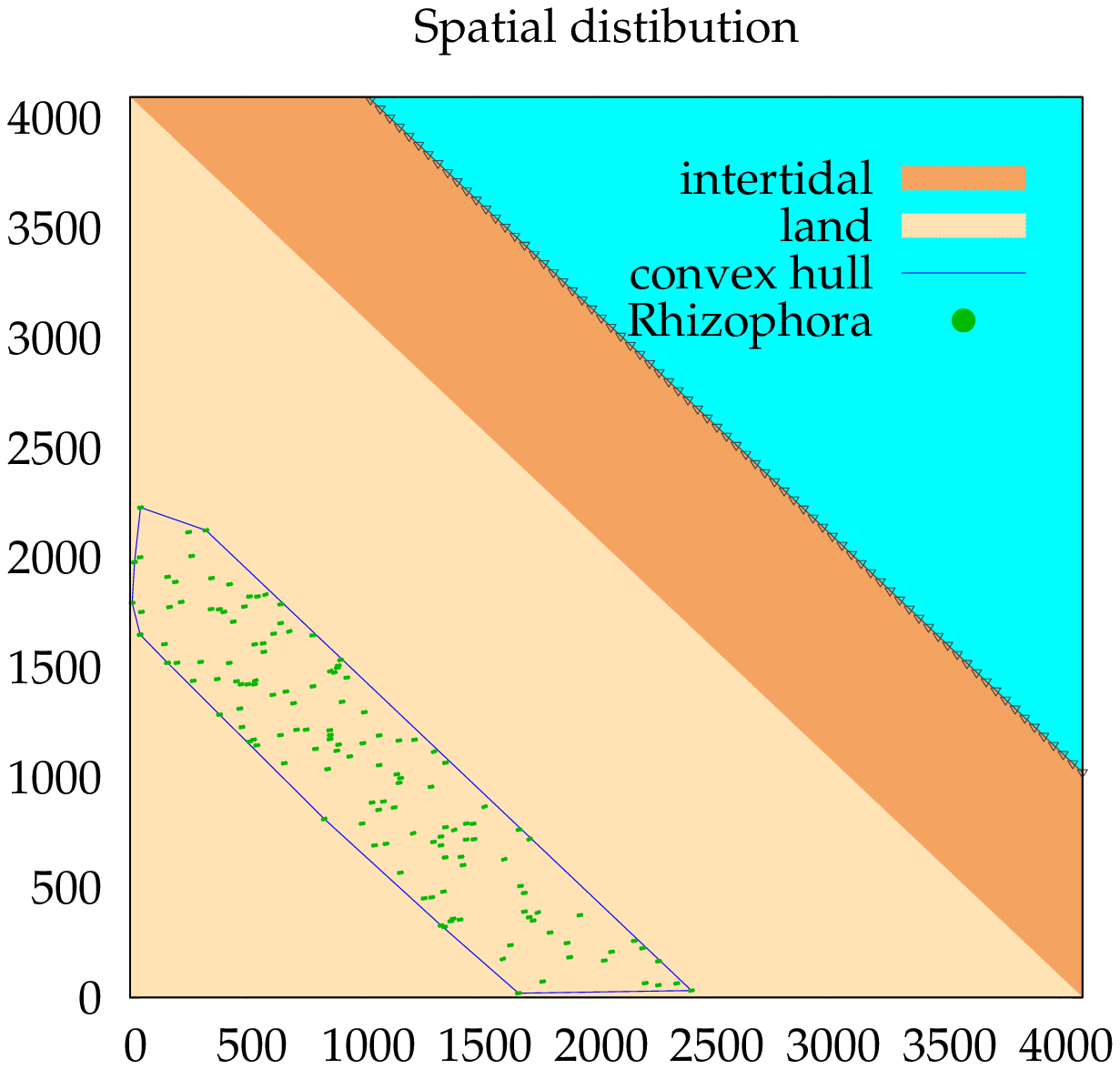}}\\
\subfloat[Arc concave-landward plot]
{\includegraphics[scale=.37]{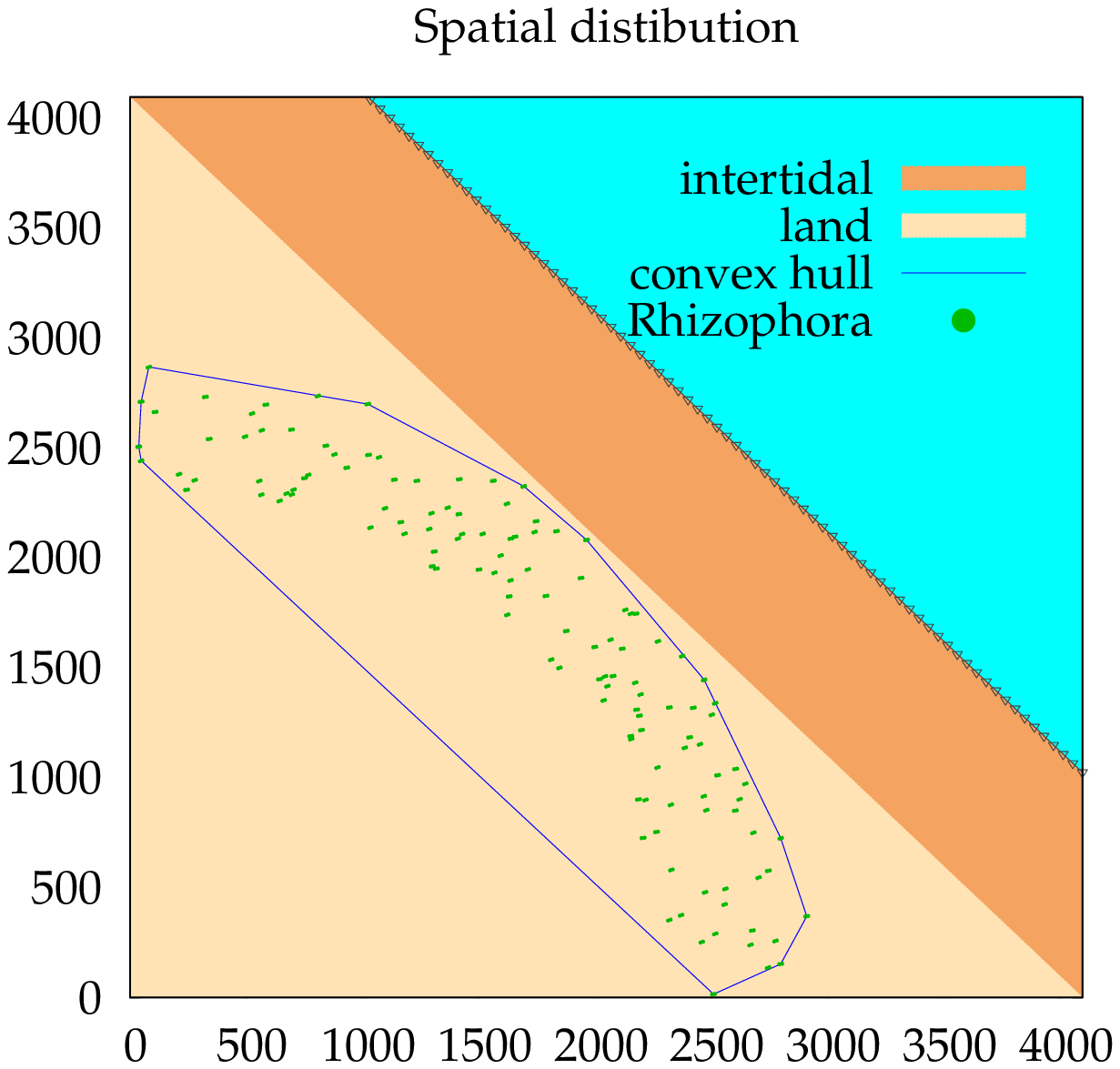}}
\subfloat[Arc concave-seaward plot]
{\includegraphics[scale=.37]{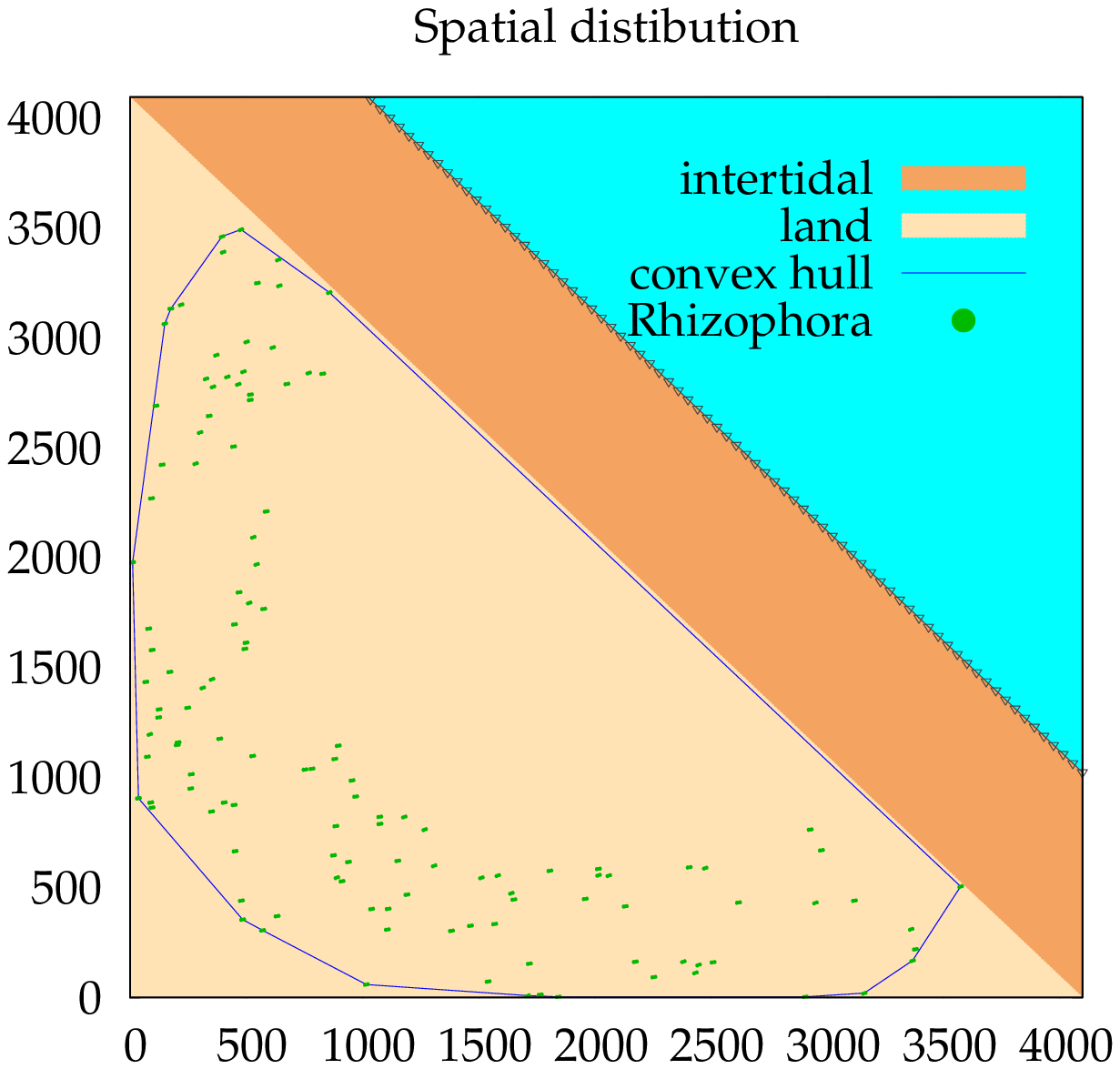}}
\caption{The configuration of the initial plot. Strip type, in order of increasing distance from the MSL line toward $(0,0)$: (a), (b), (c); Arc type: (d) concave-landward, (e) concave-seaward. For each type, the seedling density is fixed although seedling positions are randomized. The enclosing curve is the convex hull, the area $A$ of which is used to estimate densities, and the above-ground biomass per unit area.}
\label{fig:initializations}
\end{figure}

In the simulation results, the demographic rates given in Table~\ref{table:demographic-rates} are used. Two configurations of the initial plot are here considered. For configuration type, the seedling density (i.e., number of seedlings per unit area) is fixed, while the positions of individual seedlings are randomized but do not overlap. The first type of plot configuration is an approximately trapezoidal strip the long axis of which is parallel to the MSL line, as shown in Figures~\ref{fig:initializations}(a) through (c). The initial seedling density $s_d(0)$ is about $42 \mbox{ seedlings per } 100\mbox{ m}^2$. The plot partially includes the zone below MSL in Fig.~\ref{fig:initializations}(a), and the midpoints of its long sides are situated at a distance of $0.8L$ and $1.1L$ from $(0,0)$, respectively. Plots are situated totally above MSL for Fig.~\ref{fig:initializations}(b), such that the long-side midpoints are at $0.6L$ and $0.8L$ from $(0,0)$; and for~\ref{fig:initializations}(c), long-side midpoints at $0.4L$ and $0.6L$ from $(0,0)$. 

The second configuration type is an arc, which is situated totally above MSL. The seedling density $s_d(0)$ for initial plots of this type is about $13\mbox{ seedlings per } 100\mbox{ m}^2$. The arc is concave-landward in Fig.~\ref{fig:initializations}(d), and concave-seaward in Fig.~\ref{fig:initializations}(e).  

\subsection{Long-term configuration, densities, and above-ground biomass}

The performance of the forest development is presented through the seedling, sapling, and tree densities for each iteration time (time unit, $dt = 1\mbox{ day}$). By allowing virtual time to proceed through about $250\mbox{ years}$, the simulation reveals remarkably different outcomes. 

The assumptions for the simulation are as follows: human intervention (e.g., replanting, cutting down trees), and damaging environmental disturbances are absent. The above-ground biomass (AGB) is calculated using Eq.~(\ref{eq:agb}), and is also used to assess the development of the forest in time. Ideally, the tree density and AGB must be increasing in time, if not stable. A thicker forest is expected to contain more carbon stock.

\begin{figure}
\centering
\subfloat[Typical plot: $0.8L\rightarrow 1.1L$]
{\includegraphics[scale=.4]{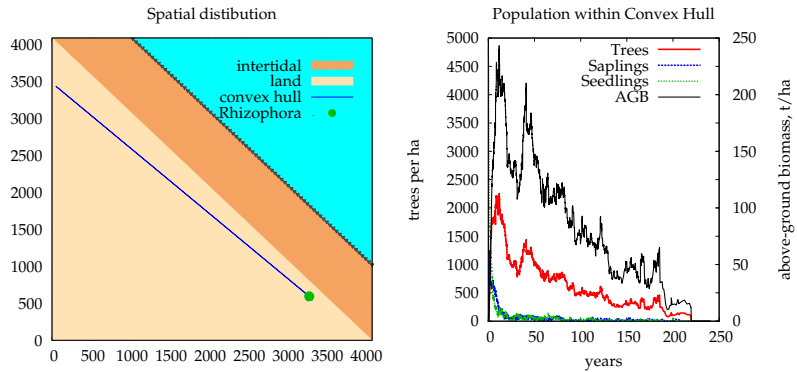}} \\
\subfloat[Above-MSL plot: $0.6L\rightarrow 0.8L$]
{\includegraphics[scale=.4]{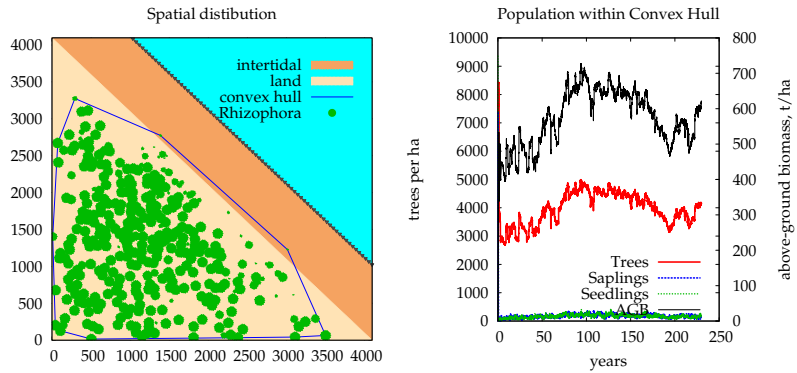}} \\
\subfloat[Above-MSL plot: $0.4L\rightarrow 0.6L$]
{\includegraphics[scale=.4]{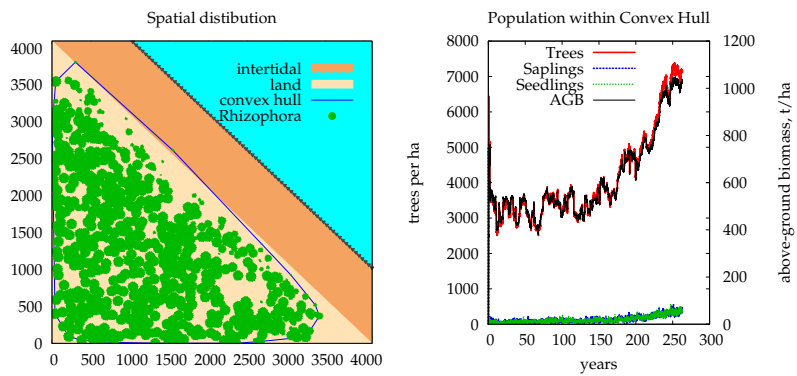}}
\caption{The long-term forest growth of plots in the order of increasing distance from the MSL line: (a), (b), and (c). For each outcome, the temporal profile of AGB, and densities for tree, sapling, and seedling are shown. Densities are values divided by the area $A$ of the convex hull. Other stochastic realizations of each configuration give rise to more or less the same outcome, even if the specific details are different.}
\label{fig:outcomes1}
\end{figure}

The typical plot, which covers zones above and below the MSL [Fig.~\ref{fig:initializations}(a)], thins out progressively until no more plants exist after about 250 years, as illustrated by Fig.~\ref{fig:outcomes1}(a). The AGB also trends down consistently along with the decrease in the tree, sapling, and seedling densities. In other words, the forest becomes extinct, although some intermittent peaks are apparent along the course of the growth trajectory. 

Above-MSL plots perform relatively better in the long run, as illustrated by Fig.~\ref{fig:outcomes1}(b) and~\ref{fig:outcomes1}(c). The observation may be explained in part by \textit{Rhizophora}'s less tolerance for salinity, the level of which is assumed to increase seaward. The mangroves are also subjected to more frequent wave action the nearer they are to the MSL line. Due to a combination of both major stressors, and a seaward gradient of both, the \textit{Rhizophora} forest is expected to thrive better at seashore locations farther away from the MSL. Mathematical explanations are provided in a later section. 

The tree densities display remarkably different trajectories among the various plots considered. Even if the area is bigger for a strip nearest the sea [Fig.~\ref{fig:initializations}(a)], the trajectory of its density and AGB are attracted towards a state of decline. In the long run, and in the absence of human intervention and weather disturbances, the mangrove forest goes extinct. Meanwhile, above-MSL plots have a better chance of survival owing to weaker stressors. 

\begin{figure}
\centering
\subfloat[Arc concave-landward plot]
{\includegraphics[scale=.4]{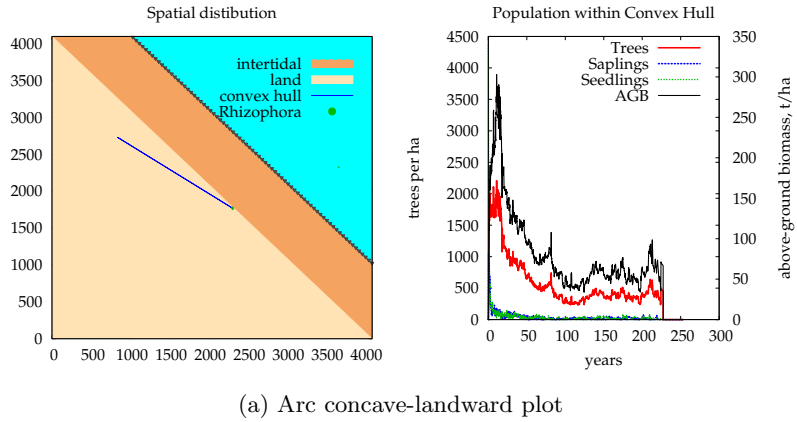}}\\
\subfloat[Arc concave-seaward plot]
{\includegraphics[scale=.4]{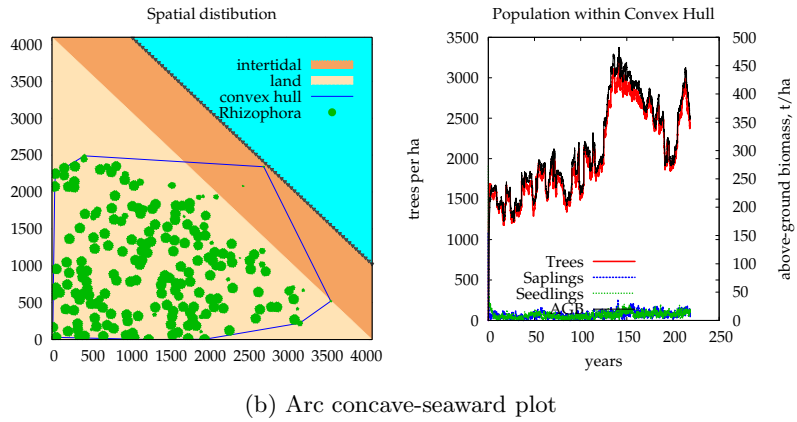}}
\caption{The long-term forest growth of arc-type plots: (a) concave-landward; (b) concave-seaward. The initial seedling densities are fixed for both settings. Other trials for the said settings yield similar results, even if the specific details can be different due to the stochasticity inherent in the model.}
\label{fig:outcomes2}
\end{figure}

The arc-type configurations depicted in Figures~\ref{fig:initializations}(d) and~(e) also induce growth trajectories which are consistent with the stressor hypothesis. The concave-landward plot includes a substantial portion just above but near the MSL where stressors are stronger. On the other hand, the concave-seaward plot has most of the seedlings situated in inland zones with weaker stressors. Consequently, the trajectory of the concave-landward plot is directed towards extinction whereas that of the concave-seaward plot is not, as illustrated by Fig.~\ref{fig:outcomes2}.

\subsection{Mathematical analysis: the bifurcation mechanism}
The dependence of the long-term performance of the restored monospecific \emph{Rhizophora} forest to the configuration of the initial plot of seedlings is ultimately related to the dynamical nature of the system. A dynamical control parameter, $\xi$, changes the nature of the system's extinction state (a fixed point). Such qualitative change is referred commonly as a bifurcation. From the linear approximation, it is evident that the only fixed point $\Gamma^*$ of the system is the state of extinction defined by the ordered density-triple: $\left(m,s_p,s_d\right) = (0,0,0) \equiv \Gamma^*$. 

The control parameter $\xi$ is implicitly a function of the spatial configuration of the system due to the stressor gradients. The initial configuration, as illustrated in Fig.~\ref{fig:initializations}, sets up the control parameter value. Here, we analyze mathematically the population dynamics of the system through its linear approximation expressed by Eq.~(\ref{eq:ODEsystem}).

First, the parameter $\xi$ is defined in terms of the pertinent rates, as follows:
\begin{equation}
\xi = \frac{k_0\tilde{D}'_5\tilde{D}'_{2.5}}{k_1k_2k_3}
\label{eq:control-parameter-xi}
\end{equation}
which acts as some sort of index the value of which is dictated by the demographic and life-stage transition rates. The spatial dependence of the parameter can be found in the product of the transition rates $\tilde{D}'_5\tilde{D}'_{2.5}$. The spatial dependence reflects the stressor gradients and the competition field.

The characteristic polynomial $p(\lambda)$ of the matrix in the linear approximation given by Eq.~(\ref{eq:ODEsystem}) is written in terms of $\lambda$ and parameterized by $\xi$ as follows:
\begin{equation}
p(\lambda) = \lambda^3 + \left(k_1+k_2+k_3\right)\lambda^2 + \left(k_1k_2+k_2k_3+k_3k_1\right)\lambda + \left(1-\xi\right)k_1k_2k_3
\label{eq:characteristic-polynomial}
\end{equation}
Given that all rates are non-negative, the last term could determine the nature of the roots of $p(\lambda)$, which correspond to the eigenvalues of Eq.~(\ref{eq:ODEsystem}). Particularly, it would depend on whether $\xi < 1$ or $\xi > 1$. In both cases $\Gamma^*$ is a hyperbolic fixed point.

For $\xi < 1$, all terms in $p(\lambda)$ are positive; but for $p(-\lambda)$ there are three sign changes from the first through the last term of Eq.~(\ref{eq:characteristic-polynomial}). Based on Descartes' sign rule, the implication of those sign changes is the existence of either three negative real eigenvalues, or one negative real eigenvalue and a pair of complex eigenvalues with negative real parts. Either case, if $\xi < 1$ then $\Gamma^*$ is a stable fixed point of the linear approximation.

On the other hand, for $\xi > 1$, the last term is the only negative term which means that there is a single sign change in $p(\lambda)$. In other words, a positive real eigenvalue is assured. But for $p(-\lambda)$, two sign changes occur through the terms in Eq.~(\ref{eq:characteristic-polynomial}) which imply the existence of either two negative real eigenvalues (in the case of a saddle point), or a pair of complex eigenvalues with negative real part (in the case of a saddle focus). Both cases nevertheless imply that $\Gamma^*$ is an unstable fixed point. 

In order to further specify whether or not the system has a pair of complex eigenvalues, the discriminant $\Delta$, which is expressed in the following equation, is evaluated.
\begin{eqnarray}\nonumber
\Delta &=& \left(k_1+k_2+k_3\right)^2\left(k_1k_2+k_2k_3+k_3k_1\right)^2 - 4\left(k_1+k_2+k_3\right)\left(k_1k_2+k_2k_3+k_3k_1\right)^3\\\nonumber
&& - 4\left(k_1+k_2+k_3\right)^3k_1k_2k_3\left(1-\xi\right) - 27k_1^2k_2^2k_3^2\left(1-\xi\right)^2\\
&&+ 18\left(k_1+k_2+k_3\right)\left(k_1k_2+k_2k_3+k_3k_1\right)k_1k_2k_3\left(1-\xi\right)
\end{eqnarray}

It turns out that with the demographic rates provided in Table~\ref{table:demographic-rates}, $\Delta < 0$ as long as $\xi$ is greater than about $0.11152$. If that is the case, then a pair of complex eigenvalues indeed exist. Since $\Gamma^*$ is hyperbolic, then the complex eigenvalues simply mean that $\Gamma^*$ is either a stable-focus node for $\xi < 1$, whereas it is a saddle-focus for $\xi > 1$. In order to further examine the implication of the above mathematical analyses, on the basis of the results presented in Fig.~\ref{fig:outcomes1} and Fig.~\ref{fig:outcomes2},  we turn our attention to the tree density as a function of time, $m(t)$. Trees represent the life stage that would most likely survive in the long run, as a type--III survivorship curve suggests. Moreover, trees carry the bulk of the AGB, as one can deduce from Figures~\ref{fig:outcomes1} and~\ref{fig:outcomes2}.

\begin{figure}
\centering
\subfloat[Typical plot: $0.8L\rightarrow 1.1L$]
{\includegraphics[scale=.4]{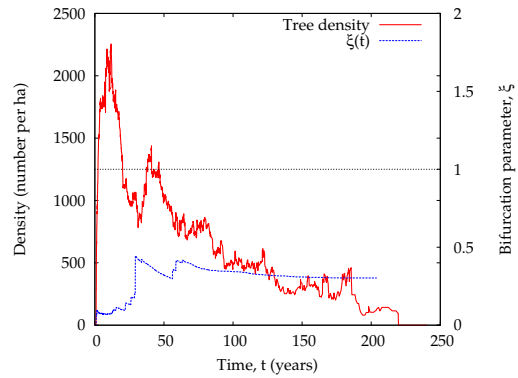}}
\\
\subfloat[Above-MSL plot: $0.6L\rightarrow 0.8L$]
{\includegraphics[scale=.4]{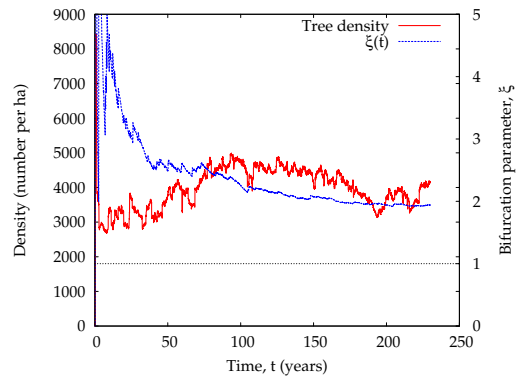}}
\\
\subfloat[Above-MSL plot: $0.4L\rightarrow 0.6L$]
{\includegraphics[scale=.4]{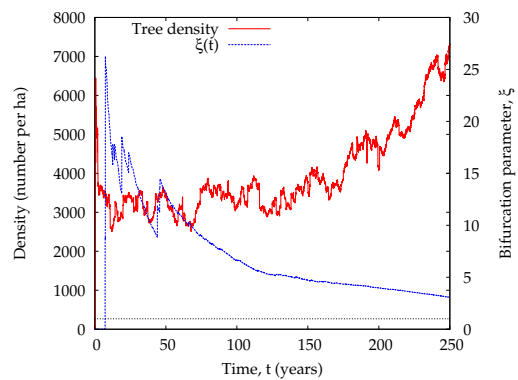}}
\caption{The bifurcation parameter $\xi(t)$ for plots with strip-type configurations superimposed with the tree density. The dashed, narrow (blue) curve represents $\xi(t)$. The horizontal line corresponds to $\xi = 1$. The solid, thick (red) curve represents $m(t)$.}
\label{fig:bifurcation1}
\end{figure}

Figure~\ref{fig:bifurcation1} shows the behavior of $\xi(t)$ superimposed with the time course of the tree density corresponding to the outcomes for the strip configurations depicted in Fig.~\ref{fig:outcomes1}. The extinction of the simulated forest as shown in Fig.~\ref{fig:outcomes1}(a) can now be justified by association with $\xi(t) < 1$ as presented in Fig.~\ref{fig:bifurcation1}(a). Looking back at the initial configuration in Fig.~\ref{fig:initializations}(a), the seedlings planted at or below MSL line are exposed to high levels of salinity and inundation stress. Thus, $\tilde{D}'_5\tilde{D}'_{2.5}$ is low by virtue of the slow growth rates afforded by high stress, as one can deduce from Eq.~(\ref{eq:dD_dt}). Consequently, $\xi < 1$. In other words, the transition rates toward the tree stage are too slow compared to the mortality rate so that the forest does not have sufficient time to flourish before all trees die out. 

For the above-MSL plot shown in Fig.~\ref{fig:initializations}(b), the tree population apparently stabilizes. Fig.~\ref{fig:bifurcation1}(b) correspondingly shows the value of $\xi(t)$ stabilizing above $1$. The value of $\xi$ starts out at a high value, which could be explained by the favorable location of the initial plot. The seedlings are found above and further away from the MSL line, wherein the stressors have less magnitude. Even as the forest spreads in spatial extent as seen in Fig.~\ref{fig:outcomes1}(b), the parameter $\xi$ remains above $1$ even if it is seemingly decreasing in monotonic fashion towards $1$. 

As the plots are placed above and further away from the MSL line, the long-term outcome and survival of the plantations considerably improve, as shown in Fig.~\ref{fig:outcomes1}(c). The forest is thicker and wider in extent, which is ideal for the maximization of carbon stock accumulation. At about $t = 150\mbox{ years}$ the tree density (and AGB) increases in breakout fashion. The result is supported by the much higher values of $\xi(t)$ relative to $1$, as depicted in Fig.~\ref{fig:bifurcation1}(c). The increasing value of $m(t)$ as time progresses is a clear manifestation of the instability of $\Gamma^*$ for $\xi > 1$. Although the sapling and seedling densities, $s_p(t)$ and $s_d(t)$, respectively, remain low, the growth rate is amply fast so that the transitions more than compensate for tree mortality. Consequently, the forest proliferates in time with a thickening tree density, which in turn guarantees a continuous supply of seedlings necessary for renewal and succession. 

\begin{figure}
\centering
\subfloat[Arc concave-landward plot]
{\includegraphics[scale=.4]{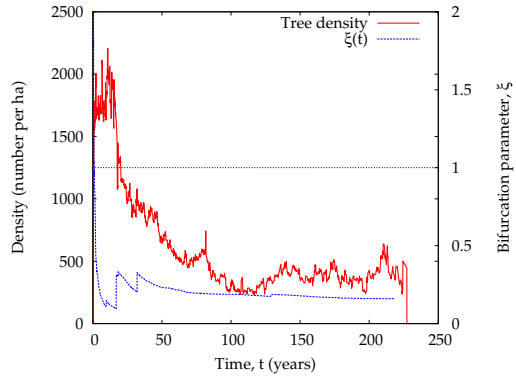}}\\
\subfloat[Arc concave-seaward plot]
{\includegraphics[scale=.4]{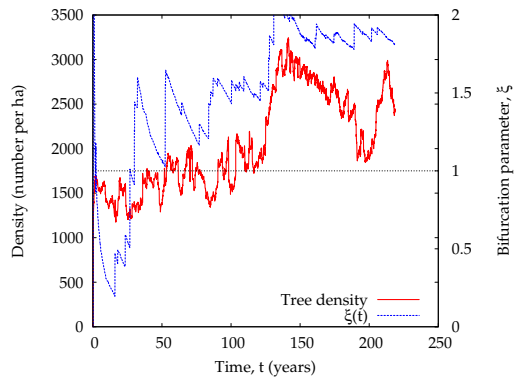}}
\caption{The bifurcation parameter $\xi(t)$ for the arc-type configurations superimposed with the tree density. The dashed, narrow (blue) curve represents $\xi(t)$. The horizontal line corresponds to $\xi = 1$. The solid, thick (red) curve represents $m(t)$.}
\label{fig:bifurcation2}
\end{figure}

A comparison of the long-term forest growth for the arc-type plot configuration based on the parameter $\xi(t)$ is presented in Fig.~\ref{fig:bifurcation2}. The concave-landward plot is associated with $\xi < 1$, as shown in Fig.~\ref{fig:bifurcation2}(a), which explains the stability of the extinction fixed point $\Gamma^*$. The parts of the plot at or below MSL are exposed to high stress levels, which result in slow growth. Hence, the transition rates $\tilde{D}'_5$ and $\tilde{D}'_{2.5}$ are too low to support a large tree density over a long period of time. This configuration emphasizes the unsuitability of positions at or below MSL in the case of \textit{Rhizophora} species used to reforest a region with seaward gradients of salinity and inundation stresses, as considered in the present study. Although some portions of the plot are situated above MSL, their population is not sufficient to increase the sample average values for the transition rates. Thus, the forest ultimately proceeds toward extinction in the long run. 

On the other hand, the concave-seaward plot has most of its seedling population above the MSL wherein stress levels are relatively weaker. Consequently, the forest is able to achieve and maintain a sufficiently high tree density over a long period of time, as illustrated in Fig.~\ref{fig:outcomes2}(b). The result is confirmed by $\xi > 1$ as time progresses, as shown in Fig.~\ref{fig:bifurcation2}(b). Although $\xi(t)$ started out below $1$, its value crossed above $1$ and stayed there since about time $t = 40\mbox{ years}$. The value of $\xi$ is somewhat only marginally above $1$, however. In some stochastic instantiations of this setting (not shown), the parameter $\xi$ fails to make the cross over to above $1$. The forest goes extinct in those cases. 

The dynamic parameter $\xi$ can be interpreted as an index for gauging the success of a particular set of initial conditions of the restored forest. In some cases, certain features of $\xi(t)$ lead the trends in the tree density. For instance, in Fig.~\ref{fig:bifurcation2}(b) the time when $\xi$ crosses from below to above the value $\xi=1$ leads the surge in the tree density by as much as $100\mbox{ years}$. In other words, we can address the question of whether or not a forest will flourish in the long run by evaluating the index within a reasonable time since restoration has commenced. With mere knowledge from field measurements of the rates of seedling establishment, plant mortality, and transition between life stages; and of stress gradients in a site of planned restoration, an estimate of the value of the index $\xi$ could be obtained within $5$ to $25\mbox{ years}$. The index $\xi$ can be used as a practical tool for assessing the long-term growth, development, and carbon stock accumulation of restored mangroves. Other possible restoration strategies as represented by their initial plot configurations may be explored in the same manner as presented in this paper.

\section{Discussion}
We have proposed, simulated, and analyzed a stochastic model for mangrove forest growth in coastal settings with seaward gradient of salinity and inundation stress. The model offers a simple description of the essential features of a coastal setting relevant to mangrove growth such as stress levels, stress gradients, and boundaries. Although we have not incorporated human factors, nor weather disturbances, the model is sufficiently generic to incorporate these factors as additional rates at relevant timescales. 

One novel aspect of our approach is the mathematical scheme of expressing the progress and transition of individual mangrove growth from seedling to sapling to tree. Instead of utilizing mathematical functions determined through regression, we instead made use of scaling relations that represent natural allometry. Not only does this method express the growth equation in terms of a single physical variable (i.e., the $\dbh$), but also it provides a truly species-specific growth model that is site-independent. Prevailing approaches are site-dependent in terms of determining the growth equation parameters and rates.

We also incorporated a stochastic scheduling procedure for determining which demographic and life-stage transition event would happen to any individual mangrove at any given point in time. The choice is ultimately dictated by random chance owing to our lack of precise and complete information characterizing the state of the system at every instant of time. The importance of stochasticity could not be further emphasized. The advantage of such approach is that we need not explicitly set fixed times during which mangroves die, or when they develop from sapling to tree. Thus, instead of setting a global clock, the Poisson process is used as a natural alternative for scheduling, which also accounts for observable randomness. Our model also considers stochasticity in representing propagule dispersal, although a more accurate description could be made by explicitly considering the influence of currents, tides, and waves especially at the intertidal zone. Lastly, by coupling the demographic processes with the spatial distribution of stress variables (which includes resource competition between individual mangroves), we have added tractability to the model that allows for insightful mathematical analysis. 

In the mathematical analysis of the linear approximation of the model, we have discovered a useful index, $\xi$, for pre-determining whether or not a certain restoration strategy would succeed in the long run in terms of promoting stable tree density, and ultimately, carbon stock accumulation. 
The index is a simple mathematical ratio between relevant rates that are measurable in the field by means of statistical sampling. The value $\xi = 1$ represents a threshold between success and failure of a restoration strategy.  The importance of an index is that we can tell within a few years whether or not a particular restoration strategy would allow a forest to actually flourish and be self-sustaining in the absence of human assistance (e.g., re-planting in forest gaps). Of course, extreme storm events could impose severe damage on restored sites, but under the guidance of a success index like $\xi$, at least one can deduce the ease by which the damage can be repaired through the self-healing ability that a stable forest possesses.

\cite{Donato} have found that the total carbon content of tropical multi-species mangrove forests is in the order of $1000\mbox{ tC/ha}$. We have found that a restored monospecific (\textit{R. mucronata}) forest associated with $\xi > 1$ could achieve half of that carbon stock in about $200$ to $250$~years post planting [see Fig.~\ref{fig:outcomes1}(c)]. On the other hand, forests restored using strategies associated with $\xi < 1$ are bound to exhibit a decreasing amount of carbon stock through the decimation of trees due to natural mortality induced by stressors and competition. Furthermore, as presented in Figures~\ref{fig:bifurcation1} and~\ref{fig:bifurcation2}, the crossing of the threshold $\xi = 1$ occurs within $5$ to $20$ years post planting. Thus, by measuring $\xi$ within that reasonable amount of time, an early indication of whether or not the restored forest shall flourish is obtained. 

Current restoration strategies in the Philippines commonly use \textit{R. mucronata} and other related species due to the relative ease by which their seedlings are tended in nurseries. Following the species zonation concept, the species from the \textit{Sonneratia} and \textit{Avicennia} genera are more appropriate to plant at the low intertidal zone~\citep{Primavera2008}. Due to the lack of appreciation for the species zonation concept and of foresight put into restoration efforts (and understanding on the ecological requirements of the chosen species), most of the restored forests are designed in such a way that \textit{Rhizophora} seedlings are planted at the low intertidal areas. However, \textit{R. mucronata} is a stenohaline species that survives better where the fluctuations of salinity are narrower and inundation frequency is low. The low intertidal, where salinity fluctuations are wide in combination with high inundation frequency, is therefore not the most suitable zone for such mangrove species. Based on our simulation results, the long-term chances for survival of restored forests designed in such a manner is between slim to null. Indeed, in the absence of a success index to guide restoration efforts, high costs of replanting are incurred without generating any real long-term positive returns. For example, in plans of building bio-shield greenbelt zones, a thick forest density [of at least $1\mbox{ km}$ width~\citep{McIvor2012}] must be guaranteed to effectively function as carbon sink and protect the coastal areas against natural disasters such as storm surges.

\section*{Acknowledgments}
This research was funded by the Commission on Higher Education (CHED) Philippine Higher Education Research Network (PHERNet). A portion of the study was presented at the PAMS12 conference in Tacloban City, Philippines in October 2013. Support from the 2013 SATU Joint Research Scheme with the International Wave Dynamics Research Center of the National Cheng Kung University, Taiwan is also acknowledged.

 \appendix







\bibliographystyle{model2-names}

\end{document}